\documentclass{revtex4}
\usepackage{amsmath}
\usepackage{amssymb}
\usepackage{amsthm}
\usepackage{amsfonts}
\usepackage{relsize}
\usepackage[colorlinks=true]{hyperref}
\usepackage{graphicx}

\begin{document}
\title{A CCCZ gate performed with 6 T gates}

\date{\today}
\author{Craig Gidney}
\email{craig.gidney@gmail.com}
\affiliation{Google Quantum AI, California, USA}

\author{N. Cody Jones}
\affiliation{Google Quantum AI, California, USA}

\begin{abstract}

We construct a CCCZ gate using six T gates, assisted by stabilizer operations and classical feedback.
More generally, we reduce the T cost of a $C^{n}Z$ gate from $4n-4$ to $4n-6$, for $n > 2$.
\end{abstract}

\maketitle

\begin{figure}
    \centering
    \resizebox{\linewidth}{!}{\includegraphics{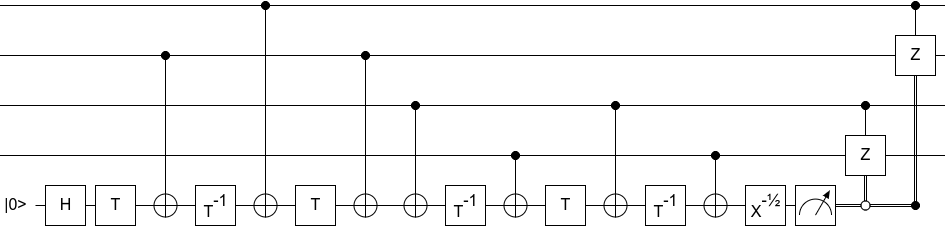}}
    \caption{A CCCZ circuit built using only 6 T gates. \href{https://algassert.com/quirk\#circuit=\%7B\%22cols\%22\%3A\%5B\%5B\%22H\%22\%2C\%22H\%22\%2C\%22H\%22\%2C\%22H\%22\%5D\%2C\%5B\%22\%E2\%80\%A2\%22\%2C1\%2C1\%2C1\%2C1\%2C\%22X\%22\%5D\%2C\%5B1\%2C\%22\%E2\%80\%A2\%22\%2C1\%2C1\%2C1\%2C1\%2C\%22X\%22\%5D\%2C\%5B1\%2C1\%2C\%22\%E2\%80\%A2\%22\%2C1\%2C1\%2C1\%2C1\%2C\%22X\%22\%5D\%2C\%5B1\%2C1\%2C1\%2C\%22\%E2\%80\%A2\%22\%2C1\%2C1\%2C1\%2C1\%2C\%22X\%22\%5D\%2C\%5B\%22~hhqv\%22\%2C\%22~hhqv\%22\%2C\%22~hhqv\%22\%2C\%22~hhqv\%22\%2C1\%2C\%22~hhqv\%22\%2C\%22~hhqv\%22\%2C\%22~hhqv\%22\%2C\%22~hhqv\%22\%5D\%2C\%5B1\%2C1\%2C1\%2C1\%2C\%22~reka\%22\%5D\%2C\%5B1\%2C1\%2C1\%2C1\%2C\%22H\%22\%5D\%2C\%5B1\%2C1\%2C1\%2C1\%2C\%22Z\%5E\%C2\%BC\%22\%5D\%2C\%5B1\%2C\%22\%E2\%80\%A2\%22\%2C1\%2C1\%2C\%22X\%22\%5D\%2C\%5B1\%2C1\%2C1\%2C1\%2C\%22Z\%5E-\%C2\%BC\%22\%5D\%2C\%5B\%22\%E2\%80\%A2\%22\%2C1\%2C1\%2C1\%2C\%22X\%22\%5D\%2C\%5B1\%2C1\%2C1\%2C1\%2C\%22Z\%5E\%C2\%BC\%22\%5D\%2C\%5B1\%2C\%22\%E2\%80\%A2\%22\%2C1\%2C1\%2C\%22X\%22\%5D\%2C\%5B1\%2C1\%2C\%22\%E2\%80\%A2\%22\%2C1\%2C\%22X\%22\%5D\%2C\%5B1\%2C1\%2C1\%2C1\%2C\%22Z\%5E-\%C2\%BC\%22\%5D\%2C\%5B1\%2C1\%2C1\%2C\%22\%E2\%80\%A2\%22\%2C\%22X\%22\%5D\%2C\%5B1\%2C1\%2C1\%2C1\%2C\%22Z\%5E\%C2\%BC\%22\%5D\%2C\%5B1\%2C1\%2C\%22\%E2\%80\%A2\%22\%2C1\%2C\%22X\%22\%5D\%2C\%5B1\%2C1\%2C1\%2C1\%2C\%22Z\%5E-\%C2\%BC\%22\%5D\%2C\%5B1\%2C1\%2C1\%2C\%22\%E2\%80\%A2\%22\%2C\%22X\%22\%5D\%2C\%5B1\%2C1\%2C1\%2C1\%2C\%22X\%5E-\%C2\%BD\%22\%5D\%2C\%5B1\%2C1\%2C1\%2C1\%2C\%22Measure\%22\%5D\%2C\%5B1\%2C1\%2C\%22\%E2\%80\%A2\%22\%2C\%22Z\%22\%2C\%22\%E2\%97\%A6\%22\%5D\%2C\%5B\%22\%E2\%80\%A2\%22\%2C\%22Z\%22\%2C1\%2C1\%2C\%22\%E2\%80\%A2\%22\%5D\%2C\%5B\%22~a48c\%22\%2C\%22~a48c\%22\%2C\%22~a48c\%22\%2C\%22~a48c\%22\%2C1\%2C\%22~a48c\%22\%2C\%22~a48c\%22\%2C\%22~a48c\%22\%2C\%22~a48c\%22\%5D\%2C\%5B\%22\%E2\%80\%A2\%22\%2C\%22\%E2\%80\%A2\%22\%2C\%22\%E2\%80\%A2\%22\%2C\%22Z\%22\%5D\%2C\%5B1\%2C1\%2C1\%2C\%22\%E2\%80\%A2\%22\%2C1\%2C1\%2C1\%2C1\%2C\%22X\%22\%5D\%2C\%5B1\%2C1\%2C\%22\%E2\%80\%A2\%22\%2C1\%2C1\%2C1\%2C1\%2C\%22X\%22\%5D\%2C\%5B1\%2C\%22\%E2\%80\%A2\%22\%2C1\%2C1\%2C1\%2C1\%2C\%22X\%22\%5D\%2C\%5B\%22\%E2\%80\%A2\%22\%2C1\%2C1\%2C1\%2C1\%2C\%22X\%22\%5D\%2C\%5B\%22H\%22\%2C\%22H\%22\%2C\%22H\%22\%2C\%22H\%22\%5D\%5D\%2C\%22gates\%22\%3A\%5B\%7B\%22id\%22\%3A\%22~a48c\%22\%2C\%22name\%22\%3A\%22verify\%22\%2C\%22matrix\%22\%3A\%22\%7B\%7B1\%2C0\%7D\%2C\%7B0\%2C1\%7D\%7D\%22\%7D\%2C\%7B\%22id\%22\%3A\%22~hhqv\%22\%2C\%22name\%22\%3A\%22test\%22\%2C\%22matrix\%22\%3A\%22\%7B\%7B1\%2C0\%7D\%2C\%7B0\%2C1\%7D\%7D\%22\%7D\%2C\%7B\%22id\%22\%3A\%22~reka\%22\%2C\%22name\%22\%3A\%22\%7C0\%3E\%22\%2C\%22matrix\%22\%3A\%22\%7B\%7B1\%2C0\%7D\%2C\%7B0\%2C1\%7D\%7D\%22\%7D\%5D\%7D}{(Click here to open in Quirk)}}
    \label{fig:circuit}
\end{figure}

This short note reports a small serendipitous discovery that occurred while I was discussing a circuit optimization problem \cite{overlaptofquestion2021} with Cody.
I was using Quirk \cite{quirk2016} to demonstrate ideas, and Cody suggested an idea for how to approach the problem based on targeting one ancilla with two Toffolis so that a pair of T gates would cancel where the Toffolis met.
I'd tried a similar idea before, and was going through the motions of showing how it didn't quite work, when Cody pointed out that a CCCZ kickback had suddenly appeared in the circuit despite there only being 6 T gates.
I isolated the CCCZ, cleaned up the circuit, and produced \hyperref[fig:circuit]{Figure \ref*{fig:circuit}}.

The circuit works because $i^{ab \oplus cd} = i^{ab} i^{cd} (-1)^{abcd}$.
The circuit uses two Toffolis to prepare $ab \oplus cd$ onto an ancilla, where $a,b,c,d$ are the computational basis values of four qubits.
The Toffolis are computed using four T gates instead of seven, to get phase kickback onto the controls.
The Toffolis touch in a way that cancels two more T gates.
After the ancilla is computed, it is phased and then removed using measurement based uncomputation.
The kickback from the Toffolis cancels the $i^{ab} i^{cd}$ components that arise from phasing the ancilla.
This leaves behind only the $(-1)^{abcd}$ term, which is the CCCZ.

A Pauli gate with $n > 2$ controls can be performed with $4n-4$ T gates by using $n-2$ temporary AND gates to cut the controls down to a pair controlling a Toffoli \cite{gidney2018halving}.
The final AND gate and the Toffoli can be replaced by a 6T CCCZ, saving 2 T gates, reducing the T cost to $4n-6$.
Oddly, we've otherwise not found any way to iterate or nest or generalize the 6T CCCZ in a way that performs common operations with lower T costs.

The 6T CCCZ circuit is simple enough that this note is likely a rediscovery, but we aren't aware of prior art.
We did find that Table I of Beverland et al. 2019 \cite{beverland2019stabilizernullity} claims that a $|CCCZ\rangle$ state can be prepared using six T gates, suggesting they knew the construction.
However, the table references a paper~\cite{jones2013} with a construction that uses eight T gates, not six.
When we contacted Beverland et al. they guessed that the T count listed in the table was most likely a typo.
Assuming that's the case, we're happy to report that the typo was merely ahead of its time.
\\

\textbf{References}

\bibliographystyle{plainnat}
\bibliography{refs}

\end{document}